# Work hardening behavior in a steel with multiple TRIP mechanisms

M.C. McGrath, D.C. Van Aken, N.I. Medvedeva, J. E. Medvedeva

ABSTRACT

Transformation induced plasticity (TRIP) behavior was studied in steel with composition Fe-0.07C-2.85Si-15.3Mn-2.4Al-0.017N that exhibited two TRIP mechanisms. The initial microstructure consisted of both ε- and α-martensites with 27% retained austenite. TRIP behavior in the first 5% strain was predominately austenite transforming to ε-martensite (Stage I), but upon saturation of Stage I, the ε-martensite transformed to α-martensite (Stage II). Alloy segregation also affected the TRIP behavior with alloy rich regions producing TRIP just prior to necking. This behavior was explained by first principle calculations that revealed aluminum significantly affected the stacking fault energy in Fe-Mn-Al-C steels by decreasing the unstable stacking fault energy and promoting easy nucleation of ε-martensite. The addition of aluminum also raised the intrinsic stacking fault energy and caused the ε-martensite to be unstable and transform to α-martensite under further deformation. The two stage TRIP behavior produced a high strain hardening exponent of 1.4 and led to ultimate tensile strength of 1165 MPa and elongation to failure of 35%.

I. INTRODUCTION

Transformation induced plasticity (TRIP) steels are expected to take a leading role in new vehicle designs that meet the 54.5 mpg corporate fuel economy average in 2025.[1] Target properties of these new steels would be combinations of ultimate tensile strengths and elongation to failures of 1000 MPa and 30% or 1500 MPa and 20%.[2] Modified TRIP steels have been reported to achieve 1000 MPa ultimate tensile strength and 30% elongation in a Fe-0.4C-1.2Mn-1.2Si steel consisting of ferrite, bainite, and retained austenite that transformed to α-martensite during deformation. Multiple martensitic transformation mechanisms have been reported where ε-martensite is sometimes reported as an intermediate phase as austenite transforms to α-martensite.[3,4] Saturation of ε-martensite in the microstructure has led to early fracture in TRIP alloys that do not subsequently transform to α-martensite.[5] It has been shown that the ε-martensite plates act as stress concentrators and limit the strength and ductility of the alloy.[6,7] Sun et al.[8] modeled the ductile failure mechanism in dual phase steels and showed that the overall ductility was primarily influenced by the mechanical strength disparity between the two phases when the martensite volume fraction was greater than 0.15. A large discrepancy between the ferrite and martensite would lead to void nucleation at lower strains, which would reduce the uniform elongation.[8] The focus of this paper was to demonstrate improved mechanical properties in a TRIP steel when ferrite was avoided and TRIP behavior to ε-martensite and α-martensite was promoted by manipulation of the generalized stacking fault energy curve.

Formation of ε-martensite is dependent upon the intrinsic stacking fault energy ($\gamma_{ISFE}$). Thermodynamic methods for calculating $\gamma_{ISFE}$ are commonly expressed as $\gamma_{ISFE} = 2\rho\Delta G^{\gamma-\varepsilon} + 2\sigma^{\gamma/\varepsilon}$, where $\Delta G^{\gamma-\varepsilon}$ is the difference between Gibbs free energy of γ-austenite and ε-martensite; ρ is the planar atomic density of the {111}; and $\sigma^{\gamma/\varepsilon}$ is the interfacial energy between γ and ε.[9,10] The Gibbs free energies of the phases are most often estimated from regular solution models for multicomponent systems.[11,12] Alloys with $\gamma_{ISFE}$ less than 20 mJ/m$^2$ have been shown to exhibit transformation induced plasticity (TRIP) behavior, where the austenite transforms to ε-martensite during deformation[11], while alloys with $\gamma_{ISFE}$ greater than 20 mJ/m$^2$ will not transform.[11] Alloys with $\gamma_{ISFE}$ between 12 and 35 mJ/m$^2$ exhibit twinning induced plasticity (TWIP) during deformation.[13]

The intrinsic stacking fault energy ($\gamma_{ISFE}$) is dependent on composition and temperature. Experimental studies have shown that carbon and aluminum will increase the $\gamma_{ISFE}$; while silicon will decrease the $\gamma_{ISFE}$. The effect of manganese is more complex. A parabolic dependence of $\gamma_{ISFE}$ on manganese concentration has been reported by several investigators.[12,14,15] Figure 1 demonstrates the relationship between $\gamma_{ISFE}$ and manganese concentration in a Fe-Mn-2.4Al-2.9Si-007C steel based on the published expression by Hirth[10] and data published by Grässel et al.[11]

Strain-induced α-martensite is reported to form at dislocation pile-ups on intersections of shear bands during deformation of a 304 stainless steel.[16-19] These shear bands may be ε-martensite, deformation twins, or densely packed stacking faults.[17-21] Olson et al.[19] modeled the kinetics for nucleation of α-martensite accounting for stacking fault energy and chemical driving force in a 304 stainless steel. Intrinsic stacking fault energy decreased with decreasing temperature, which increased the rate that shear bands formed with respect to strain. The probability that a shear band intersection will generate an α-martensite embryo was found to be dependent upon the chemical driving force for the transformation.[19] The chemical driving force for α-martensite transformation is between -1090 and -1210 J/mole in the Fe-C system[22] and -1300 J/mole in Fe-30Ni steel.[23] The driving force for α-martensite, $\Delta G^{\gamma-\alpha}$, and ε-martensite, $\Delta G^{\gamma-\varepsilon}$, dictated the stable martensite phase during deformation. The growth of α-martensite occurs by repeated nucleation of embryos and coalescence.[24] Talonen et al.[25] studied the formation of strain-induced martensite in austenitic stainless steels and noted α-martensite was always observed subsequently following ε-martensite formation. These authors[25] suggested that the density of shear bands preceding α-martensite formation primarily dictated the transformation to α-martensite and the variation of the chemical driving force to α-martensite had minimal influence.

The different mechanisms for austenite transformation during deformation was reported and characterized in various studies. Yang et al.[5] produced a TRIP steel without aluminum additions (Fe-0.24C-21.5Mn) and showed austenite transformed only to ε-martensite (γ-austenite→ε); there was no evidence of α-martensite production. The stacking fault energy was calculated as 3.4 mJ/m$^2$ using published data by Grässel et al.[11] Conversely Frommeyer et al.[26] used results from interrupted tensile tests to show transformation of both austenite and ε-martensite to α-martensite (γ-austenite+ε→α) during deformation of a TRIP steel with composition Fe-15.8Mn-2.9Al-3Si-0.02C. The increased aluminum concentration in the steel tested by Frommeyer et al.[26] had an intrinsic stacking fault energy of 20 mJ/m$^2$ as a result of a high aluminum content. It should be noted, however, that the starting microstructure was duplex (free ferrite and austenite) prior to forming martensite and the composition of the austenite was not given. Cai et al.[27] produced a TRIP steel with composition Fe-0.18C-0.67Si-1.38Mn-0.56Al. The low aluminum concentration caused the stacking fault energy to decrease to 15.6 mJ/m$^2$. The microstructure prior to deformation consisted of retained austenite, bainite, and free ferrite. The austenite transformed to α-martensite during deformation without an intermediate transformation to ε-martensite. Again, the duplex microstructure and lack of chemical analysis for the austenite makes it difficult to rationalize the stacking fault energies. However, the same calculated stacking fault energy was achieved in an alloy formulated for this paper with a higher aluminum and manganese composition of Fe-0.07C-2.85Si-15.3Mn-2.4Al-0.017N. The microstructure prior to tensile testing of this alloy consists of a combination of retained austenite, ε-martensite, and α-martensite. Thermodynamic calculations of intrinsic stacking fault energy may not provide a complete picture in these TRIP materials, since the unstable stacking fault energy controls the nucleation of Shockley partial dislocations and thus the nucleation of ε-martensite.

## II. GENERALIZED STACKING FAULT CURVES FROM FIRST PRINCIPLES

First-principles atomistic modeling of stacking fault defects is instructive for understanding the deformation behavior in metals and alloys. The generalized stacking fault energy ($\gamma_{GSFE}$) describes the energies for shearing atomic planes and determines the structure and mobility of dislocations. In fcc Fe, a slip <112>(111) and a planar shift at the Burgers vector $\mathbf{b}_p$=1/6<112> produces a minimum on the GSF curve (Figure 2a) and represents the intrinsic stacking fault energy ($\gamma_{ISFE}$), where a negative $\gamma_{ISFE}$ would favor formation of the hexagonal closed-pack crystal structure (i.e. ε-martensite). The unstable stacking fault ($\gamma_{USFE}$) which corresponds to the displacement at 1/12<112>(0.5|$b_p$|), controls the barrier for the formation of stacking faults and nucleation of Shockley partial dislocations, where a reduction in $\gamma_{USFE}$ would lower the barrier for formation of both ε-martensite and deformation twins. It should be noted that the stacking fault energy values obtained with ab-initio calculation are typically magnitudes different from experimental values and thermodynamic estimates. The differences may be related to magnetic interactions and light interstitial impurities. Nonetheless, the calculations account for interatomic interactions and structure caused by alloying and provide an in-depth understanding of the mechanisms for the formation of stacking faults.

Additions of aluminum, manganese (greater than 13 at.%), and carbon were shown to increase the $\gamma_{ISFE}$. Significantly, Medvedeva et al.[28] showed that the $\gamma_{ISFE}$ depended on the location of the manganese relative to the stacking fault where manganese located one interlayer distance from the stacking fault affected the intrinsic stacking fault energy and produced a parabolic dependence of stacking fault energy with the concentration of manganese (Figure 2b) and is in excellent agreement with the results of Figure 1. In contrast, when manganese is distributed uniformly and at greater than 10 at% concentrations, the $\gamma_{ISFE}$ does not change (see Figure 2b). Also shown in Figure 2b is that the $\gamma_{USFE}$ decreases with manganese concentration. A lower energy barrier for the formation of stacking faults in the regions of high manganese concentrations should lead to the formation of ε-martensite or mechanical twins.

Medvedeva et al.[28] showed that carbon preferred to segregate away from the stacking fault and would increase the $\gamma_{ISFE}$, whereas manganese was found to increase the $\gamma_{ISFE}$ but preferred to be near the stacking fault. Mn-C defect pairs were favorable in austenite along with the 180° Mn-C-Mn complex.[28] When Mn-C-Fe and Mn-C-Mn defects were considered at the stacking fault, the $\gamma_{ISFE}$ was lower compared to the value of Fe-C and the calculated values for the stacking fault compared better with the experimental stacking fault measurements.

Aluminum additions were shown to increase the $\gamma_{ISFE}$ and decrease the energy barrier ($\gamma_{USFE}$) for Shockley partial dislocation nucleation and therefore aluminum would be expected to promote the formation of ε-martensite or deformation twins. Short range order is generally reported in high aluminum and manganese alloys. Medvedeva et al.[28] showed short range order inhibits the increase in $\gamma_{ISFE}$ while decreasing the $\gamma_{USFE}$. The addition of aluminum would be expected to lower the $\gamma_{USFE}$ which would promote the formation of ε-martensite while at the same time increase the $\gamma_{ISFE}$ which would cause ε-martensite to be less stable (Figure 2c).

The goal of this paper is to demonstrate improved mechanical properties in steel with multiple TRIP mechanisms and show that a steel can be formulated by careful manipulation of the unstable and intrinsic stacking fault energies.

III. EXPERIMENTAL PROCEDURE

A high aluminum and manganese steel was produced to study the martensitic transformation during deformation. An alloy composition was modeled after the TRIP steel that was produced by Frommeyer et al.[26] with the aim to have a starting microstructure that was free of ferrite. The carbon and nitrogen concentrations were raised and the aluminum concentration was lowered slightly to produce an alloy composition in weight percent of Fe-0.07C-2.85Si-15.3Mn-2.4Al-0.017N and with calculated intrinsic stacking fault energy of 15.9 mJ/m$^2$. High purity induction iron, electrolytic manganese, aluminum, ferrosilicon, and carbon were melted in a 45 kg induction furnace under an argon protective atmosphere. Plates were cast into phenolic no-bake olivine sand molds designed for 12.6 cm x 6 cm x 1.7 cm blocks. Chemical analysis was performed by ion coupled plasma spectrometry after sample dissolution in perchloric acid. The casting was homogenized at 1373 K (1100 °C) for 2 hours before being air cooled to room temperature. The casting was then milled to 13.6 mm x 126 mm x 50 mm. Hot rolling occurred incrementally, starting at 1173 K (900 °C) with reheating between reductions when the temperature fell below 973 K (700 °C). The plates were reduced 80% to a thickness of 2.8 mm. After the final rolling pass the plates were reheated to 1173 K (900 °C) for 10 minutes before being water quenched.

Tensile test specimens were machined from the hot rolled products in accordance to ASTM E8-08[29] with a gage section of 50 mm in length and 12.5 mm in width. Tensile tests were performed with the load axis parallel to the rolling direction. Tests were conducted at room temperature using a displacement rate of 0.01 mm/s. Interrupted tensile tests were performed where samples were loaded to various displacements, unloaded, and then metallographically prepared for x-ray diffraction measurement. A Scintag 2000 diffractometer using CuKα radiation was used to characterize strained samples cut from the gage section. Scans were run from 30-100 degrees with a scan step size of 0.03 degrees. The volume fractions of the phases were determined using the integrated intensity of the diffraction peaks and applying the equation reported by De et al.[30] Secondary electron and light optical microscopy was used to characterize the microstructures. Standard metallographic practices were employed to polish specimens and each was etched with 2% nital solution and subsequently etched with 10% sodium metabisulfate to contrast the differences between martensite and retained austenite.

IV. RESULTS

The microstructure of the TRIP alloy before tensile testing is shown in Figure 3. The hot rolled microstructure was a combination of martensite (both ε and α) and retained austenite, which was representative for the steel. The alloy contained 60% ε-martensite, 27% austenite, and 13% α-martensite prior to the tensile tests. No ferrite was observed in the microstructures.

A representative and complete curve for the stress-strain behavior of the Fe-0.07C-2.85Si-15.3Mn-2.4Al-0.017N steel is shown in Figure 4. The ultimate tensile strength was 1165 MPa at a necking strain of 0.33. The strain hardening exponent, n ($d(\ln\sigma)/d(\ln\varepsilon)$), was plotted as a function of true strain and is shown in Figure 5a. Logarithmic smoothing was applied to the curve. Between 2% and 5% the strain hardening exponent was between 0.2 and 0.4. The strain hardening exponent increased at a rapid rate from 0.4 to 1.3 with increased strain between 5% and 10% strain. The strain hardening exponent diminished from 1.4 at 15% strain to 0.6 for strains up to 26%. The strain hardening rate was derived by differentiating the true stress-true strain curve ($d\sigma/d\varepsilon$) and the result after logarithmic smoothing is plotted as a function of true strain in Figure 5b. The strain hardening rate initially decreased and reached a minimum at 5% strain. Subsequent straining produced a maximum strain hardening rate of approximately 8065 MPa at 15% strain, which corresponds to the maximum in the strain hardening exponent. At strains greater than 15% both the exponent and the rate of hardening diminished.

Figure 6 is a series of diffraction patterns obtained from interrupted tensile tests where the amount of ε-martensite and α-martensite were investigated and quantified (see Table I) with respect to applied strain. Initially, the amount of ε-martensite increased from 60% to 65% for strains less than 4%. The percent of α-martensite did not change in a statistically significant way at these low strains. At strains greater than 5% the amount of α-martensite began to increase from 15% to 45%, which correlates with the rapid increase in strain hardening exponent and strain hardening rate. The fraction of ε-martensite and austenite remained constant at strain levels between 10% and 14% and this coincided with strain hardening exponents remaining relatively constant. The amount of ε-martensite and austenite decreased at strain levels between 15% and 23% and this corresponded to the strain hardening exponents decreasing in Figure 5. After tensile failure the only phase detected by x-ray diffraction was α-martensite. Figure 7 shows microstructures of the steel transformed at different strain levels. ε-martensite was less apparent at higher strain levels.

Figure 8 shows that the martensitic transformation was not homogeneous throughout the Fe-0.07C-2.85Si-15.3Mn-2.4Al-0.017N specimen strained to 10%. SEM images of the microstructure revealed the following regions: (1) untransformed austenite; (2) austenite segmented by ε-martensite where a small fraction of α-martensite nucleated at intersections of the ε-martensite plates; and (3) segmented austenite with a large fraction of α-martensite formation at intersections of ε-martensite plates. Qualitative chemical analyses within the different regions were obtained using standardless energy dispersive spectroscopy (EDS). The EDS results from the regions described above are summarized in Table II. A higher manganese concentration was observed within the austenitic region (region 1) whereas the lowest manganese concentration was observed in the region containing α-martensite (region 3). The aluminum and silicon concentration were indistinguishable statistically. Thermodynamic stacking fault energies were calculated using the qualitative composition values to reflect a trend within the different regions and are shown in Table II. As expected, the stacking fault energies increased as manganese concentrations increased in the untransformed austenite.

V. DISCUSSION

The Fe-0.07C-2.85Si-15.3Mn-2.4Al-0.017N steel produced an ultimate tensile strength of 1165 MPa with an elongation to failure of 35%, which met targeted property goals considered break-through for a 3$^{rd}$ generation advanced high strength steel. Unlike previous studies of high manganese and aluminum TRIP steels the Fe-0.07C-2.85Si-15.3Mn-2.4Al-0.017N steel was fully austenitic prior to formation of 60% ε-martensite and 13% α-martensite. Ferrite was absent and ε-martensite dominated the starting microstructure. In contrast the Frommeyer et al.[26] steel was alloyed with more aluminum and less carbon and nitrogen to yield a composition of Fe-15.8Mn-2.9Al-3Si-0.02C that produced a duplex TRIP steel containing 37% free ferrite, 15% ε-martensite, and retained austenite. It should be noted α-martensite was not reported in the microstructure prior to tensile testing. Yang et al.[5] produced an austenitic alloy without aluminum additions but higher manganese additions (Fe-21.51Mn-0.24C) that transformed to ε-martensite. Reducing manganese and aluminum was done by Cai et al.[27] to produce a composition of Fe-0.18C-0.67Si-1.38Mn-0.56Al. This steel had a microstructure consisting of free ferrite, bainite, and retained austenite that transformed to α-martensite upon deformation.

The TRIP character of our steel was similar to the 304 stainless steels.[3,4] Upon application of tensile strain and for up to 5% strain, the steel reported here transformed to ε-martensite. The strain hardening rate (Figure 5b) reached a minimum as the ability of the austenite to transform to ε-martensite was exhausted during Stage I of transformation. Further strain produced very rapid hardening rates as both austenite and ε-martensite were transformed to α-martensite during Stage II. It is uncertain whether the austenite transformed directly to α-martensite or transformed first to ε-martensite. It could be suggested that ε-martensite must form first based upon the low strain (<5%) transformation behavior where only ε-martensite formed. Between 10% and 14% strain the proportions of the phases are relatively constant indicating no phase transformation occurred. Figure 9a shows untransformed austenitic regions in the steel deformed to 10% strain. Inclusions present in these regions suggest this region was the last to solidify and thus would have a higher alloy concentration and this was confirmed by qualitative chemical analyses to be higher in manganese (see Table II). Solute partitioning during solidification was taken into account by using a method of Scheil segregation modeling with FactSAGE[31] and a composition of Fe-22.3Mn-5Si-1.3Al-0.25C was predicted in the last 10% of the metal to solidify. These regions of higher manganese concentration stabilized the austenite and produced a higher calculated intrinsic stacking fault energy (see Table II). A higher stacking fault energy is expected to require a higher stress for transformation[25,32] and thus would delay the transformation to higher strains. Figure 9b shows the segregated regions with high density of inclusions were transformed after 20% strain. The segregated regions were segmented as ε-martensite formed (Stage I) after 20% strain. The α-martensite etched at a faster than the austenite or the ε-martensite and Figure 9c shows ε-martensite transformed to α-martensite in segregated regions with a high inclusion density after 20% strain indicating the region to solidify last. X-ray diffraction shows that α-martensite increased rapidly at strains greater than 15% and the metallographic results suggest that the nature of the transformation did not change. In these segregated regions ε-martensite still appears to form first and provide nucleation sites for α-martensite. The oscillatory variation in the strain hardening rates at strains greater than 10% (see Figure 5b) may be a reflection of chemistry variation (i.e. segregation) and the strain required to initiate the TRIP in these different regions.

The maximum strain hardening exponent observed was 1.4 which was approximately 1.5 times larger than the maximum strain hardening exponent observed by Frommeyer et al.[26] Differences in the initial amount of ε-martensite may explain the difference in strain hardening behavior. It has been reported that ε-martensite provides barriers to the movement of slip dislocations and causes dislocation pile-ups resulting in high strain hardening exponents.[6,7] In contrast, the TRIP steel produced by Cai et al.[27] showed constant strain hardening during plastic deformation (n~ 0.3) and transformed only to α-martensite. It would thus appear that a large initial proportion of ε-martensite is required to obtain the high strain hardening rates. However, lower quantities of retained austenite may limit ductility. The TRIP alloy produced for this study achieved an elongation of 35% (true strain ~30%), which was less than that reported by Frommeyer et al.[26] In comparison to Frommeyer et al.[26] the starting amount of austenite was less (27% vs. 48%) and the austenite was completely transformed at failure. Frommeyer et al.[26] concluded that continuous martensitic transformation was responsible for achieving an elongation to failure of 45% where the volume fraction of retained austenite decreased from 48% to 15%.

Chemical driving forces dictate the probability for α-martensite formation. The chemical driving forces for ε-martensite and α-martensite transformation for the various TRIP compositions discussed in this paper were calculated and are compared in Table III. The chemical driving force for α-martensite plates at room temperature has been related to the α-martensite start temperature,[33,34] where the α-martensite start temperature equation accounted for manganese, carbon, and silicon;[35] the effect of aluminum on martensite start temperature is still uncertain.[36-38] A regular solution model determined by Grässel et al.[11] was used to calculate the driving force for ε-martensite formation. The chemical driving forces for α-martensite from ε-martensite in the steels produced for this study and by Frommeyer et al.[26] confirmed ε-martensite would be less stable relative to α-martensite, which explains the transformation to α-martensite during deformation. The last region to solidify in the present steel was predicted to have a lower driving force to ε-martensite than the bulk composition and was not as prone to transform as observed in the higher solute concentrated regions. A consequence of suppressing ε-martensite formation at low strains was less nucleation sites for α-martensite, which delayed the martensitic transformation in the steel. In contrast the driving force for α-martensite from ε-martensite was calculated for the steel produced by Yang et al.[5] as 75 J/mole, which confirms ε-martensite was more stable than α-martensite. As a result, ε-martensite did not transform to α-martensite during deformation even though there was high saturation of ε-martensite present that could act as nucleation sites for α-martensite. The hypothesis that strain-induced α-martensite was primarily controlled by the stacking fault energy and chemical driving force had minimal influence on α-martensite formation[25] was not supported with the observations in the TRIP alloy produced by Yang et al.[5] It is interesting to note that the driving force for austenite to ε-martensite in the steel produce by Cai et al.[27] was equivalent to the present steel even though Cai et al.[27] did not report ε-martensite formation. Thus the stacking fault energy predicted by the thermodynamic model does not appear to accurately predict ε-martensite formation in low manganese and aluminum steels. First principles calculations of stacking fault energy accounted for structure and interatomic interactions caused by alloying; whereas the thermodynamic predictions accounts for the driving force to ε-martensite as a function of composition and temperature. First principles calculations also showed that aluminum lowered the unstable stacking fault energy ($\gamma_{USFE}$), which would promote ε-martensite formation. This may explain why the TRIP steel studied by Cai et al.[27] did not form ε-martensite. Aluminum was also shown to increase the intrinsic stacking fault energy ($\gamma_{ISFE}$) and this would encourage ε-martensite transformation to α-martensite. Cai et al.[27] alloyed with much lower amounts of manganese and aluminum. In low manganese steels the manganese was shown to be uniformly distributed and would have less effect on the $\gamma_{USFE}$ but lower the $\gamma_{ISFE}$. Thus, it may be hypothesized that two stage TRIP is obtained with alloying additions that must lower the $\gamma_{USFE}$ to obtain ε-martensite while the additions should also raise the $\gamma_{ISFE}$ to encourage α-martensite formation. This effect was possible in the very high manganese and aluminum steel formulated in this paper.

VI. CONCLUSION

Transformation induced plasticity (TRIP) behavior was studied in steel with composition Fe-0.07C-2.85Si-15.3Mn-2.4Al-0.017N that exhibited a maximum work hardening exponent of 1.4. The work hardening behavior was outstanding due to the high fraction of ε-martensite formed. The high strain hardening rate led to an ultimate strength of 1165 MPa at a necking strain of 33% and transformation of austenite to martensite (γ-austenite→ε→α) led to enhanced elongation. Segregation in the steel led to regions that were expected to have a higher stacking fault energy and thus delayed transformation to higher strains. The delay in transformation resulted in oscillatory variation in the strain hardening rates at strains greater than 10% as the segregated regions transformed.

The composition of the TRIP steel was formulated based on calculations for stacking fault energy and chemical driving forces for the ε-martensite and α-martensite transformation from austenite. Chemical driving forces for the two variants of martensites were compared and agreed with microstructures reported in several TRIP steels. Thermodynamic stacking fault energy predictions failed at considering martensite transformation in low manganese and aluminum compositions. First principles calculations of stacking fault energy gave a deeper understanding of the formation of ε-martensite with respect to alloying elements. Addition of aluminum decreased the unstable stacking fault energy, which promoted ε-martensite formation. Aluminum was also shown to increase the intrinsic stacking fault energy, which caused the ε-martensite to be unstable and transform to α-martensite under further deformation.

ACKNOWLEDGEMENTS

This work was supported in part by the National Science Foundation (NSF) and the Department of Energy under contract CMMI 0726888. The authors gratefully acknowledge the support of the Graduate Center for Materials Research and in particular Dr. Eric Bohannen for help with x-ray diffraction. Meghan McGrath was supported by a Department of Education GAANN fellowship under contract P200A0900048.

LIST OF FIGURES

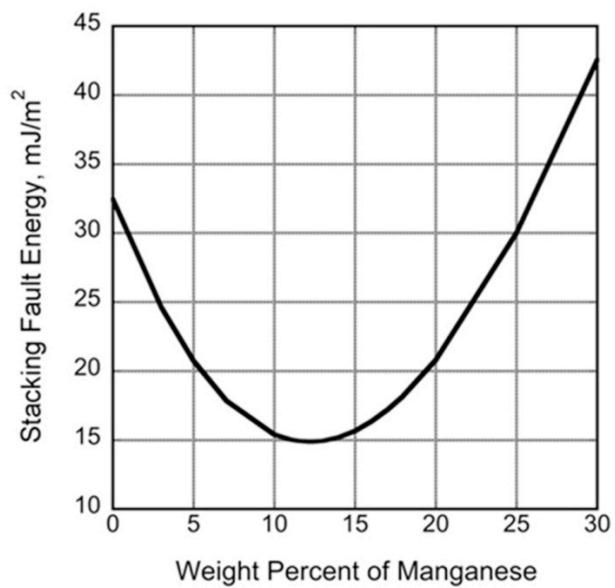

Fig. 1 – Model of stacking fault energy for Fe-Mn-2.4Al-2.9Si-0.07C with varying manganese content. The minimum stacking fault energy occurs at approximately 12.25 wt.% manganese. The model was based on published expressions by Hirth[10] and thermodynamic data published by Grässel et al.[11]

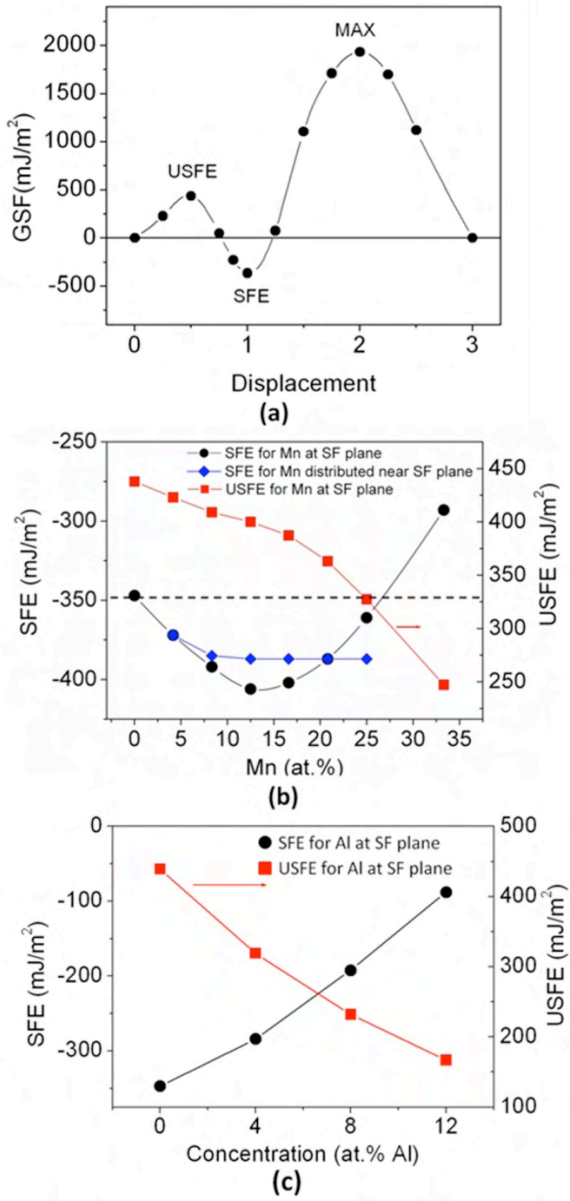

Fig. 2 – (a) Example of the generalized stacking fault (GSF) energies for FCC Fe. The intrinsic stacking fault energies (SFE) and unstable stacking fault energies (USFE) as function of (b) manganese and (c) aluminum concentration.[28]

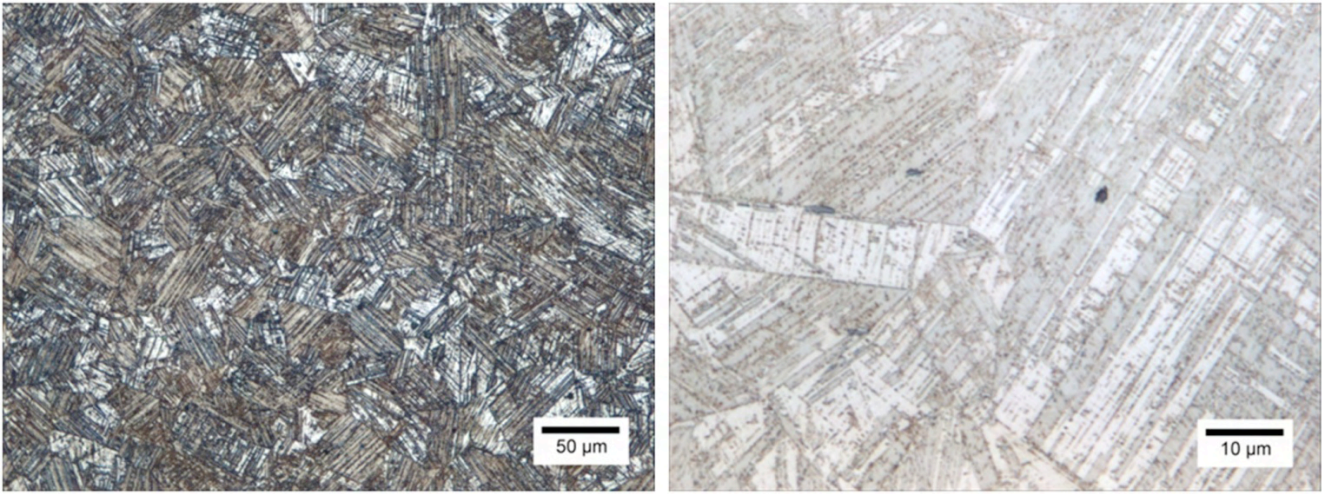

Fig. 3 – Short-longitudinal microstructures of Fe-0.07C-2.85Si-15.3Mn-2.4Al-0.017N before tensile testing. The prior austenite grain size was measured to be 29 ± 2.6 (95% confidence level) by the Heyn intercept method.

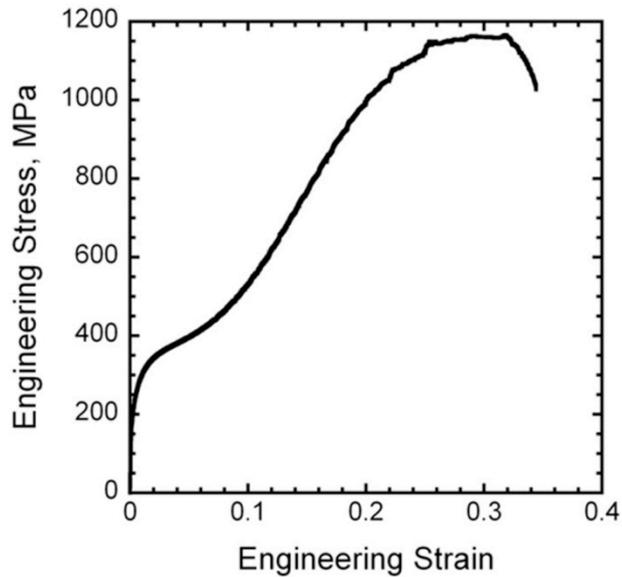

Fig. 4 – Representative tensile curve of Fe-0.07C-2.85Si-15.3Mn-2.4Al-0.017N after hot rolling to an 80% reduction, reheated to 1173 K (900 °C) for 10 minutes, and quenched in water to ambient temperature.

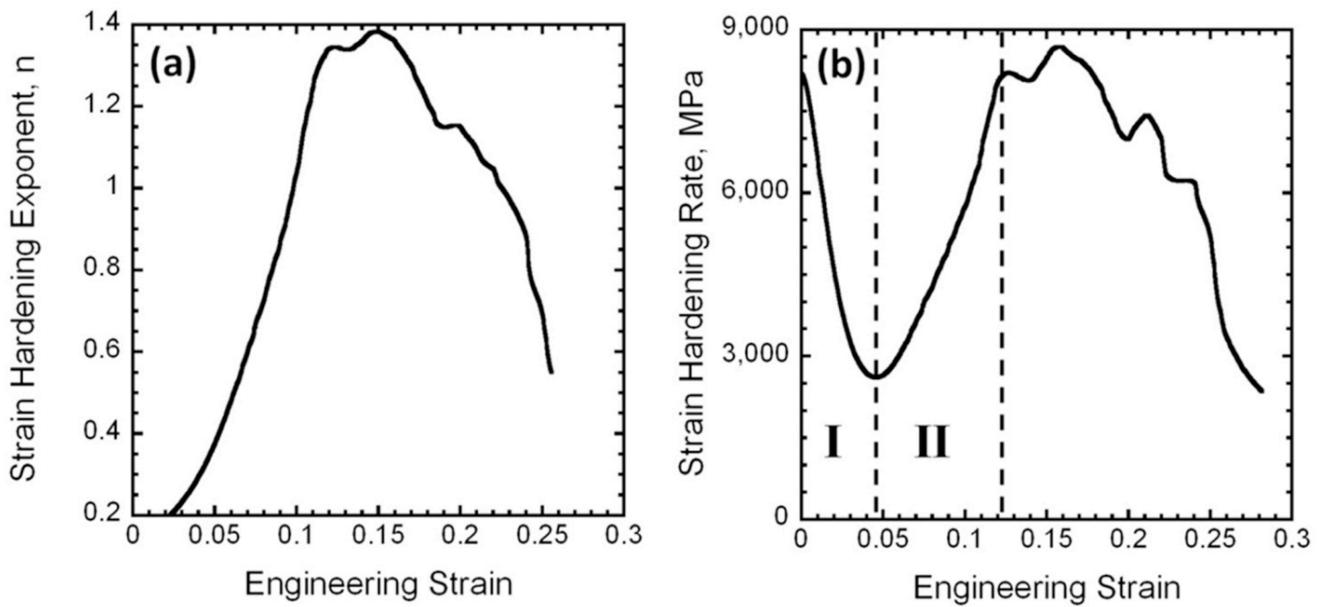

Fig. 5 – (a) Strain hardening exponent and (b) strain hardening rate varied as a function of engineering strain.

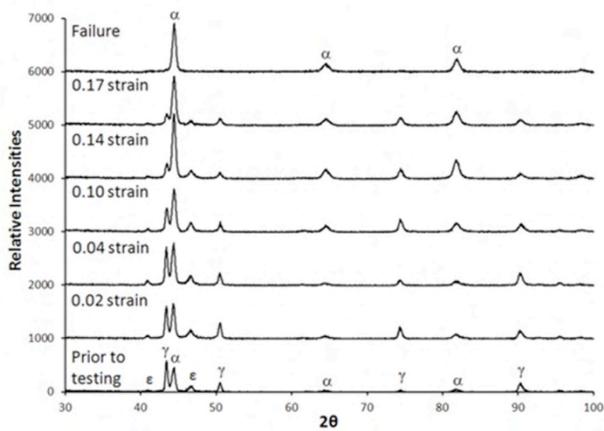

Fig. 6 – X-ray diffraction patterns of Fe-0.07C-2.85Si-15.3Mn-2.4Al-0.017N at various strains during tensile testing. ε-martensite peaks could be observed in all patterns up to failure.

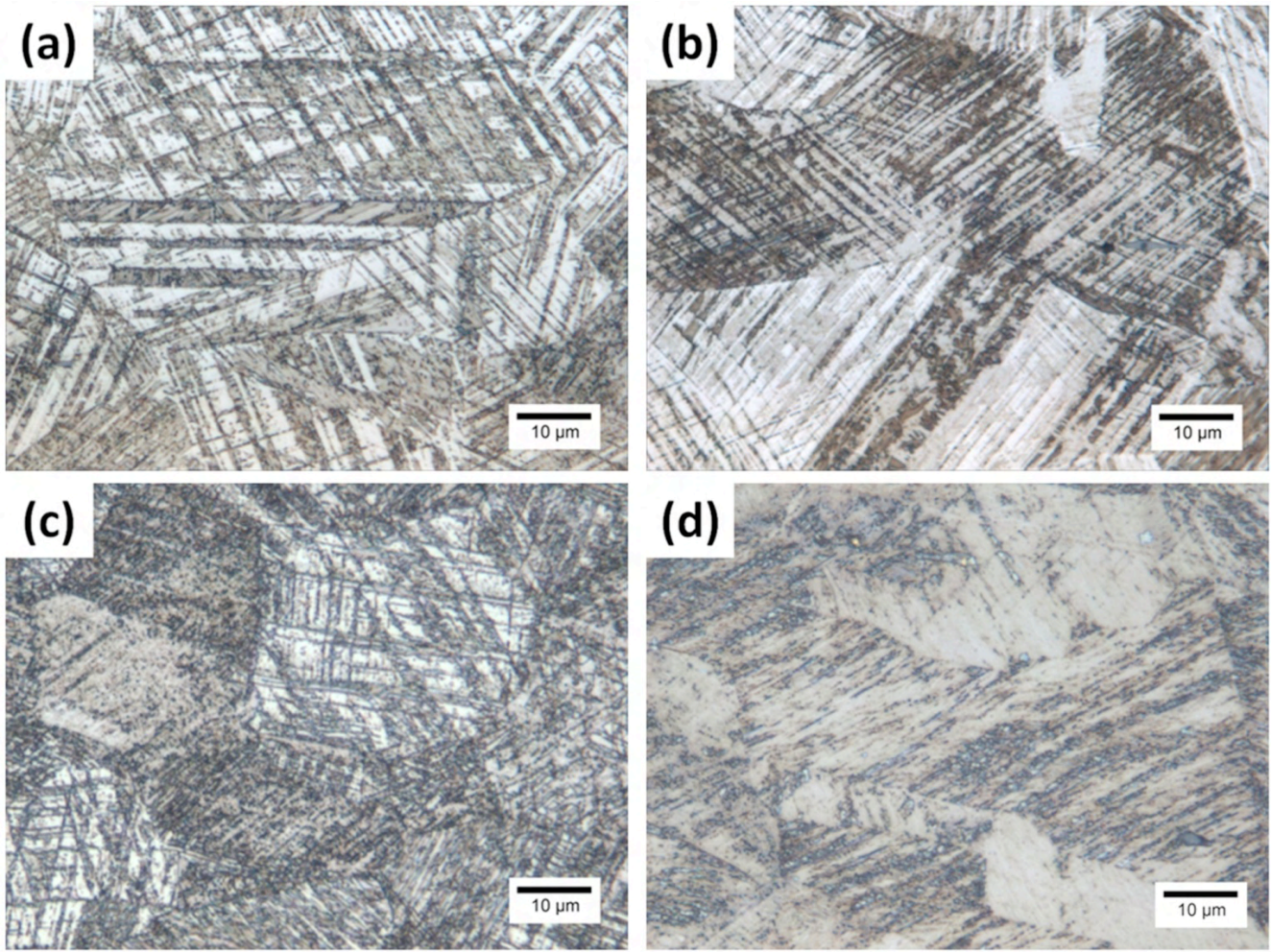

Fig. 7 – Short-longitudinal microstructures of Fe-0.07C-2.85Si-15.3Mn-2.4Al-0.017N at (a) 2%, (b) 10%, (c) 17%, and (d) 35% strain levels during tensile testing. The steel was hot rolled and quenched in water from 1173 K (900 °C) prior to testing.

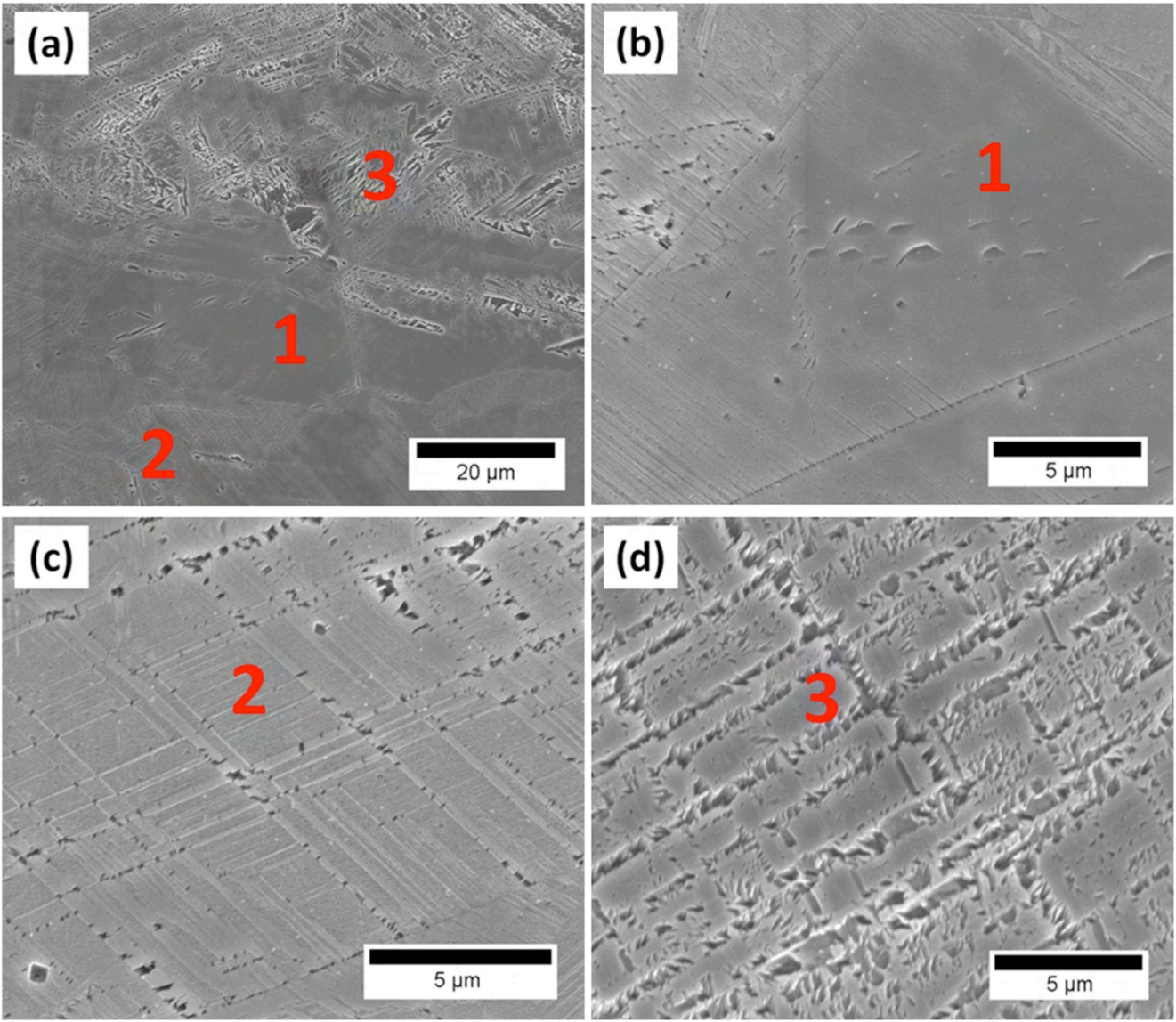

Fig. 8 – Secondary electron image of α-martensite within ε-martensite Fe-0.07C-2.85Si-15.3Mn-2.4Al-0.017N at 10% strain.

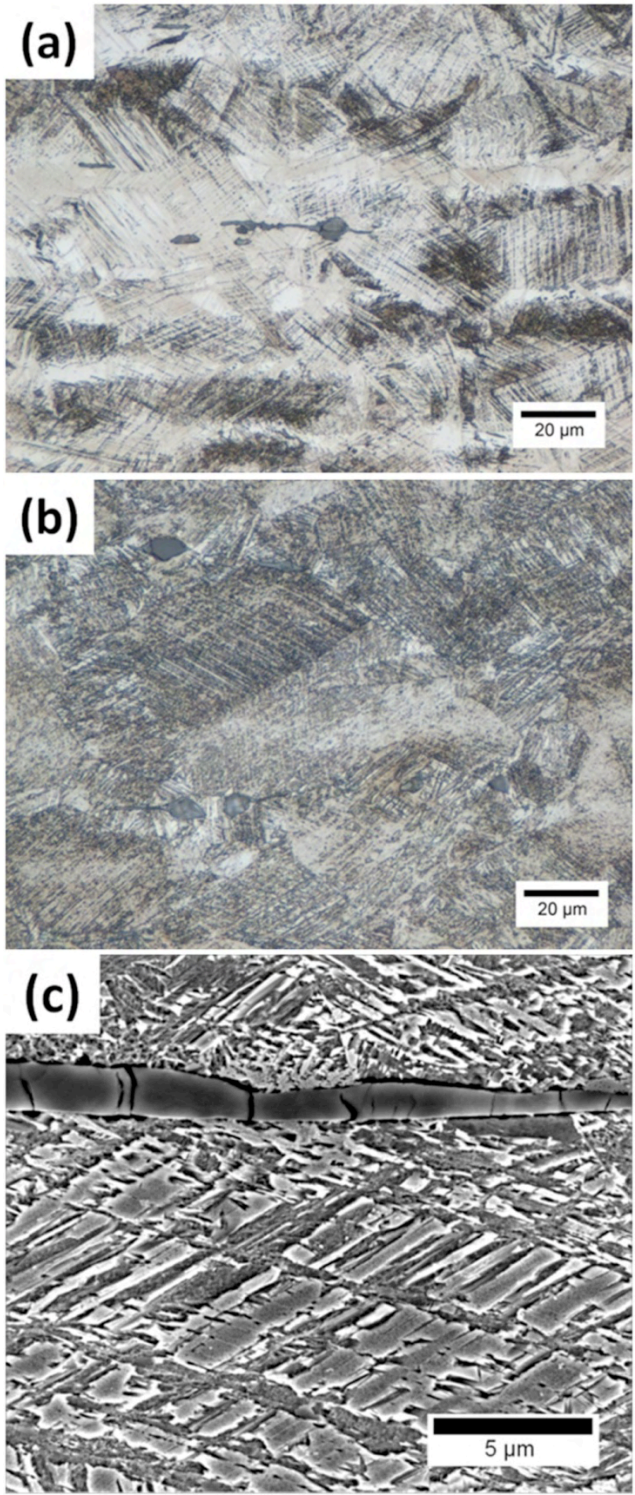

Fig. 9 – Solute segregation led to untransformed austenite in a sample strained to (a) 10%. Segregated regions were transformed in samples strained to (b) 20%. (c) Secondary electron image showing the TRIP behavior at 20% strain in a Mn-segregated region.

LIST OF TABLES

Table I.  Volume Fraction of Phases Present at Various Strains

| Strain | γ-austenite | ε-martensite | α-martensite |
|--------|-------------|--------------|--------------|
| 0.00 | 0.268 | 0.598 | 0.134 |
| 0.02 | 0.228 | 0.612 | 0.160 |
| 0.04 | 0.199 | 0.651 | 0.150 |
| 0.10 | 0.103 | 0.450 | 0.447 |
| 0.14 | 0.097 | 0.447 | 0.456 |
| 0.17 | 0.065 | 0.331 | 0.604 |
| 0.23 | 0.056 | 0.191 | 0.752 |
| 0.34 | 0.000 | 0.000 | 1.000 |

Table II. Summary of Chemical Analysis and Calculated Thermodynamic Stacking Fault Energy in Various Regions of a Specimen Strained 10%

| Region | Al, wt.% | Si, wt.% | Mn, wt.% | $\gamma_{ISFE}$, mJ/m$^2$ |
|--------|----------|----------|----------|--------------------------|
| 1 | 1.20 ± 0.14 | 2.49 ± 0.13 | 18.82 ± 0.75 | 8.0 |
| 2 | 1.17 ± 0.12 | 2.23 ± 0.21 | 17.46 ± 0.79 | 6.3 |
| 3 | 1.67 ± 0.44 | 1.96 ± 0.32 | 15.09 ± 0.74 | 4.3 |

Table III.  Calculated Driving Forces for ε- and α- Martensite

| Composition used from Reference: | $\Delta G^{\gamma \to \varepsilon}$, J/mole | $\Delta G^{\gamma \to \alpha}$, J/mole | $\Delta G^{\varepsilon \to \alpha}$, J/mole |
|---|---|---|---|
| Present TRIP | -84 | -862 | -778 |
| Last Region to Solidify Predicted with Scheil Model | -77 | -90 | -13 |
| Frommeyer et al. [26] | 8 | -925 | -933 |
| Yang et al. [5] | -338 | -263 | 75 |
| Cai et al. [27] | -88 | -2036 | -1948 |